\documentclass[12pt]{article}

\newcommand{\bra}[1]{\langle{#1}|}
\newcommand{\ket}[1]{|{#1}\rangle}
\newcommand{\braket}[2]{\langle{#1}|{#2}\rangle}

\newcommand{\eq}[1]{Eq.~(\ref{#1})}

\def\beq{\begin{equation}}
\def\eeq{\end{equation}}
\def\beqa{\begin{eqnarray}}
\def\eeqa{\end{eqnarray}}

\newcommand{\EQ}{\begin{equation}}
\newcommand{\EN}{\end{equation}}
\newcommand{\bea}{\begin{eqnarray}}
\newcommand{\ena}{\end{eqnarray}}

\renewcommand{\a}{\alpha}
\renewcommand{\b}{\beta}
\renewcommand{\c}{\gamma}

\newcommand{\shalf}{\frac{1}{2}}

\newcommand{\NP}[1]{Nucl.\ Phys.\ {\bf #1}}
\newcommand{\PL}[1]{Phys.\ Lett.\ {\bf #1}}

\def\one{{\hbox{ 1\kern-.8mm l}}}
\def\gh{{\rm gh}}
\def\sgh{{\rm sgh}}
\def\NS{{\rm NS}}
\def\R{{\rm R}}
\def\ii{{\rm i}}

\def\comm#1#2{\left[ #1, #2\right]}
\def\acomm#1#2{\left\{ #1, #2\right\}}
\def\tr{{\rm tr\,}}
\newcommand{\N}{{\cal N}}
\newlength{\bredde}
\def\slash#1{\settowidth{\bredde}{$#1$}\ifmmode\,\raisebox{.15ex}{/}
\hspace*{-\bredde} #1\else$\,\raisebox{.15ex}{/}\hspace*{-\bredde} #1$\fi}

\begin{document}\begin{titlepage}
\begin{flushright} KUL-TF-98/31 \\ hep-th/9808074
\end{flushright}
\vfill
\begin{center} 
{\LARGE\bf Anomalous D-brane and orientifold
\vskip 2.mm
 couplings from the boundary state}    \\
\vskip 27.mm  \large
{\bf   Ben Craps $^{1,2}$, Frederik Roose $^3$ } \\
\vskip 1cm
{\em Instituut voor theoretische fysica}\\
{\em Katholieke Universiteit Leuven, B-3001 Leuven, Belgium}
\end{center}
\vfill

\begin{center}
{\bf ABSTRACT}
\end{center}
\begin{quote}
We compute scattering amplitudes involving both R-R and  NS-NS fields in the presence of a
D-brane or orientifold plane. These provide direct evidence for the anomalous couplings 
in the D-brane and orientifold actions. The D9-brane and O9-plane are found to couple 
to the first Pontrjagin class with the expected relative strength.

\vfill      \hrule width 5.cm
\vskip 2.mm
{\small
\noindent $^1$ Aspirant FWO, Belgium }\\
{\small
\noindent $^2$ E-mail: Ben.Craps@fys.kuleuven.ac.be }\\
{\small
\noindent $^3$ E-mail: Frederik.Roose@fys.kuleuven.ac.be }
\end{quote}
\begin{flushleft}
\end{flushleft}
\end{titlepage}

\section{Introduction}
Since their introduction in string theory D-branes have attracted an ever increasing
interest. Accordingly the effective action describing the low energy dynamics acquired various
additional terms. Among these the Wess-Zumino couplings to the RR-potentials appear to be the
least
well-studied. Apart from the natural coupling to the $(p+1)$--form there are also interaction
terms coupling a D$p$-brane to e.g. the $(p-1)$-- and $(p-3)$--forms, for $p$ high enough. In
\cite{GHM} the precise form of these was derived from anomaly arguments\footnote{ For
explanation of the used symbols and conventions see Appendix A. }:
\beq
\label{WZaction}
S_{WZ}=\frac{T_p}{\kappa}\int_{p+1}\hat{{\cal C}'}\wedge e^{2\pi\a '\,F+\hat{B'}}\wedge
\sqrt{\hat{A}(\hat R)}~~.
\eeq
The $ p_1(\hat R)$ term in the expansion was already proposed in \cite{BerVaf} based on 
an independent duality reasoning. 
In Ref. \cite{Das} a similar coupling was proposed for orientifold planes. Although void of world
brane dynamics, 
these planes do couple to bulk fields for anomalies to cancel along the lines of Ref. \cite{GHM}.
In this
paper we will give additional evidence for some of these terms by evaluating the corresponding
string diagrams.\footnote{The $ p_1(\hat R)$ terms in Refs \cite{BerVaf,Das} actually came 
with the proposal to do the corresponding three point string amplitudes. }

Scattering amplitudes in the presence of D-branes have hitherto been evaluated in two ways;
people have either essentially performed a disk or annulus calculation \cite{klebanov,gm} 
or relied 
on the boundary state formalism \cite{billo9802,dv9707}. 
We opt for the last possibility although 
in principle the calculations could be done analogously to \cite{gm}. A nice feature of our choice
is that the results thus obtained carry over to the orientifold case almost without effort.
 
In section \ref{ss:BC} we set up the basics of doing string tree level scattering involving the
boundary state. As a working example we derive the $\int B \wedge {\cal C}_{p-1}$ coupling. 
In section \ref{ss:RRC} we outline
the procedure to find the desired $\int p_1 \wedge {\cal C}_{p-3} $ term, 
thereby relying on the basic results in the
preceding
section. Making contact with the effective D-brane action is done in
section \ref{ss:FT}. 
In section \ref{ss:ORI} we derive the tree level amplitude for the gravitational coupling
of an orientifold plane.
Section \ref{ss:CONC} then concludes and raises a few
questions concerning the origin of the studied interaction terms. An appendix explaining our
conventions and normalizations is added.

\section{The $B \wedge {\cal C}_{p-1}$ interaction}
\label{ss:BC}
Various string amplitude calculations \cite{gm, dv9707} 
established the coupling of a D$p$-brane to the
RR $(p-1)$--form potential: 
\beq 
2\sqrt{2}\pi\alpha ' T_p \,\int\;{\cal C}_{p-1}\wedge F~~,
\eeq
where $\kappa^2 = 8\pi G_N $.
From the fact that only the $2\pi\alpha ' F + B'$ combination is gauge invariant one can then
infer the presence of 
\beq
\sqrt{2} T_p\int\;{\cal C}_{p-1}\wedge \hat B'~~.
\eeq
This D-brane coupling to the NS-NS two form should reproduce the leading 
behaviour at low energies of the following
string amplitude \footnote{with polarizations chosen along the brane worldvolume.}:
\beq
\label{BC:ampl}
{\cal A} = \bra{{\cal C}_{p-1};k_2} V_B^{(0)}(\zeta,k) \ket{B}_\R~~, 
\eeq
which is equivalent with the two point function on the disk of
the Kalb-Ramond field $B$ and the RR-potential ${\cal C}_{p-1}$.
Here and in the following we adopt both the RR-state $\bra{{\cal C}_{p-1}}$
and the boundary state $\ket{B}_{\R}$\footnote{For the
precise form of the RR state and the boundary state see also Appendix A.}
from \cite{billo9802} (and we give the RR potential a momentum $k_2$.). 
In the following we carry out the
evaluation of this simple-looking amplitude using the boundary state. Thus we will find out
about the basic manipulations involved when computing any boundary state amplitude. 

First, as the boundary state and the RR potential already saturate the superghost anomaly
\cite{billo9802}
any additional vertex must be in the $(0,0)$-picture. Therefore we may take (see \eq{KR})
\beq
V_B^{(0)}=\frac{\kappa}{\sqrt{2}\pi}\,\zeta_{\mu\nu}\,\int d^2 z (\partial X^{\mu}+\ii\,k\cdot
\psi\,\psi^{\mu})(\bar\partial
X^{\nu}+\ii\,k\cdot\tilde\psi\,\tilde\psi^{\nu})~~,
\eeq
where $z$ and $\bar z$ run over the complement of the unit disk in the complex plane.

A substantial amount of work in computing amplitudes involving the boundary state 
goes into pulling the non-zero mode operators 
\beq
\exp\biggl[-\sum_{n=1}^\infty \frac{1}{n}\,
\a_{-n}\cdot S\cdot
\tilde \a_{-n}\biggr]\ ; 
\eeq
\beq \label{expferm}
\exp\biggl[\ii\eta\sum_{m=1}^\infty
\psi_{-m}\cdot S \cdot \tilde \psi_{-m}\biggr]
\eeq
from the boundary state to the left where they annihilate the out vacuum. 
For amplitudes in the RR sector one then has to take care of the fermion zero
modes. 
From \eq{convgamma} in the Appendix  
it is clear that those zero modes are effectively replaced by $\gamma$-matrices.
In
the conventions we are using any amplitude in the RR sector contains a factor 
\beq \label{factor}
\left(\frac{1}{\sqrt{2}}\right)^n \tr \left( C^{-1}{\cal M} R\,C^{-1} {\cal N}^T L \right)~~,
\eeq
where $L$ is a product of $\gamma$-matrices corresponding to the left-over left-moving
fermi\-ons, $R$ a product of $\gamma$'s for the right-moving fermions and $n$ is the total 
number of $\gamma$'s in $LR$. The matrices ${\cal M} $
and ${\cal N}$ show up in the boundary state and the RR state respectively. 
Note that \eq{factor} is
schematic: in fact, keeping track of all factors of $\gamma_{11}$ and 
minus-signs is crucial to obtain correct results.  

Equivalently \cite{klebanov}, one modifies the Green functions to include
left-right contractions:
\begin{eqnarray}
\langle X^{\mu}(z)\,\bar X^{\nu}(\bar w)\rangle&=&-S^{\mu\nu}\log
(1-\frac{1}{z\bar w})\\
\langle\psi^{\mu}(z)\,\bar\psi^{\nu}(\bar w)\rangle_R&=&\frac{S^{\mu\nu}}{2}
\frac{\ii\eta}{z\bar w-1}\frac{1+z\bar w}{\sqrt{z\bar w}}\ .
\end{eqnarray}

Before applying the above rules to the amplitude \eq{BC:ampl}, 
let us remark that the ghost sector is discussed in the Appendix,
whereas the superghost sector contributes a factor $1/2$ \cite{billo9802}.

Contracting two fermions using the massive oscillators in the boundary state\footnote{The 
zero-mode contribution is proportional to $\frac{k\cdot S\cdot k}{k\cdot k_2}$ for small on-shell 
momenta. Further investigations seem to suggest that this vanishes whenever the 
$B \wedge {\cal C}_{p-1}$ interaction could contribute to cross-sections, as we argue in
the next paragraph. We thank Rodolfo Russo for pointing out this problem to us.}, 
leads to 
\beqa
{\cal A}&=&\frac{T_p}{2}\frac{\kappa}{\sqrt{2}\pi}\frac{1}{2\sqrt{2}(p-1)!}\frac{32}{4}\shalf
\epsilon^{\a_1\ldots\a_{p-1}\b_1\b_2} c_{\a_1\ldots\a_{p-1}}\zeta_{\b_1\b_2}\nonumber\\
&&\int_{|z|>1}d^2z\,
\left(|z|^2\right)^{-k\cdot S \cdot k-k\cdot k_2}
\left(|z|^2-1\right)^{k\cdot S \cdot k}
\frac{(k\cdot S\cdot k)}{|z|^2 \left(|z|^2-1\right)}~~, 
\eeqa
where $c_{\a_1\ldots\a_{p-1}}$ is the polarization tensor of the RR field.
Doing the $\int d^2 z$ integral gives 
$\pi\, (k\cdot S \cdot k)\, B(1+k\cdot k_2,\,k\cdot S \cdot k)$.
We thus conclude that, for small momenta,
\beq \label{AmpCB}
{\cal A} = \frac{\kappa T_p}{2\,(p-1)!} \epsilon^{\a_1\ldots\a_{p-1}\b_1\b_2} 
c_{\a_1\ldots\a_{p-1}}\zeta_{\b_1\b_2}~~.
\eeq

We end this section with a remark which is related to the previous footnote. Because of the
decoupling of longitudinal polarizations, the amplitude vanishes if $k$ (or equivalently $k'$)
has a non-zero component along the brane directions\footnote{This was pointed out by Igor
Pesando.}. In computing cross-sections, one has to
average over neighbouring momenta. So in order to have a non-vanishing cross-section, one needs
the possible momentum components along the brane to be discrete, i.e. all worldvolume directions
should be compactified. This suggests considering Euclidean branes. In that case one can have
on-shell momenta without a component along the brane worldvolume. For such momenta the numerator
of the expression in the footnote vanishes. 

\section{The $\tr R^2\wedge {\cal C}_{p-3}$ interaction}
\label{ss:RRC}
We have to calculate the following amplitude:
\beq
{\cal A}=\bra{{\cal C}_{p-3};k_2}V_g^{(0)}(\zeta_3,k_3)\,V_g^{(0)}(\zeta_4,k_4)\ket{B}_{\R}~~.
\eeq
Since we are looking for the $\tr R^2\wedge {\cal C}_{p-3}$ term in the D-brane effective action, 
we will consider the components of ${\cal C}_{p-3}$ with indices along the brane. 
This implies that the fermion
zero modes have to provide the four gamma-matrices with Lorentz indices 
in the worldvolume directions complementary to the ones of
the RR potential that is being considered.

As the computation is analogous to the one presented in the previous section but considerably
longer, we only give a rough sketch. Since each graviton vertex operator can be written as the sum
of four different parts, the amplitude splits into various pieces. Some integrations by parts
in the relative angle are needed to let the various terms combine nicely. 
Performing the trace of gamma-matrices, taking into account the superghost sector and the
GSO-projection, one ends up with the following result (possibly up to a global sign):
\beqa
{\cal A}&=&
\frac{\kappa^2\,T_p\a '^2}{4\sqrt{2}\,(p-3)!\,\pi^2}\,
\epsilon^{\a_1\cdots\a_{p-3}\b_1\cdots\b_4}\,c_{\a_1\cdots\a_{p-3}}\,
k_{3\b_1}\,k_{4\b_3}\nonumber\\ 
&&[(k_4\cdot S\cdot\zeta_{3\b_2})(k_3\cdot\zeta_{4\b_4})-(k_3\cdot S\cdot k_4)(\zeta_{3\b_2}
\cdot\zeta_{4\b_4})
+(k_4\cdot\zeta_{3\b_2})(k_3\cdot S\cdot\zeta_{4\b_4})\nonumber \\ 
&&-(k_3\cdot k_4)(\zeta_{3\b_2}\cdot S\cdot\zeta_{4\b_4})]\nonumber\\
&&\int_{|z_3|,|z_4|>1}d^2z_3\,d^2z_4\,
(|z_3|^2-1)^{k_3\cdot S\cdot k_3}\,(|z_4|^2-1)^{k_4\cdot S\cdot k_4}\,
|z_3|^{-2\,k_3\cdot S\cdot k_3-2\,k_3\cdot S\cdot k_4-2} \nonumber \\ &&
|z_4|^{-2\,k_4\cdot S\cdot k_4-2\,k_3\cdot S\cdot k_4-2}\, 
|z_3-z_4|^{2\,k_3\cdot k_4-2}\,|z_3\bar z_4-1|^{2\,k_3\cdot S\cdot k_4-2}\,
(z_3\bar z_4-\bar z_3z_4)^2~~.\nonumber
\eeqa

This is the general result. However, when we want to talk about the 
worldvolume action of a D-brane,
we might be especially interested in momenta and polarizations which are along the brane 
worldvolume.
From now on we specialize to that case. Because of the mass
shell condition and momentum conservation in the worldvolume directions
the integral above simplifies to
\beq \label{Int}
{\cal I}=\int_{|z_3|,|z_4|>1}d^2z_3\,d^2z_4\,|z_3z_4|^{-2}\,
|z_3-z_4|^{-2}\,|z_3\bar z_4-1|^{-2}\,(z_3\bar z_4-\bar z_3z_4)^2~~.
\eeq

The integral depends only on the difference of the phases of $z_3$ and $z_4$, so one of the angular
integrations just gives a factor of $2\pi$. By changing variables to the cosine of the difference
of the phases, it can be verified that the second angular integration results in a hypergeometric
function. With $r=|z_3|$, $t=|z_4|$, $x=\frac{4rt}{(r+t)^2}$ and
$y=\frac{4rt}{(rt+1)^2}$ the integral becomes (up to a minus sign) \cite{grad}:
\beq
{\cal I}=\pi^2\,\int_{r,t>1}\,dr\,dt\,(rt)^{-1}\,xy\,F_1(\frac{3}{2},1,1,3;x,y)~~.
\eeq
Performing a change of integration variables and manipulating the hypergeometric function, this
becomes
\beq
{\cal I}=\frac{\pi^2}{2}\,\int_0^1\,dy\,(1-y)^{-1/2}\,
\sum_{n=0}^{\infty}\frac{(\frac{3}{2})_n}{(3)_n}\,y^n\,\int_0^1\,dx\,(1-x)^{-1/2}
F(\frac{3}{2}+n,1,3+n;x)~~.
\eeq
In this expression, both integrals are well-known \cite{grad}. The result is\footnote{We 
thank Walter Troost for showing us how to sum the following series.}
\beq \label{Intresult}
{\cal I}=\frac{\pi^3}{2}\,\sum_{k,n=0}^{\infty}\,
\frac{(\frac{3}{2})_{n+k}\,\Gamma(k+1)\,\Gamma(n+1)}{(3)_{n+k}\,
\Gamma(k+\frac{3}{2})\,\Gamma(n+\frac{3}{2})}=\frac{4\,\pi^4}{3}~~.
\eeq

The scattering amplitude thus becomes
\beq \label{AmpCRR}
{\cal A}=
\frac{\sqrt{2}\,\pi^2\,\kappa^2\,T_p\a '^2}{3\,(p-3)!}\,
\epsilon^{\a_1\cdots\a_{p-3}\b_1\cdots\b_4}\,c_{\a_1\cdots\a_{p-3}}\,k_{3\b_1}\,k_{4\b_3}
(k_4\cdot\zeta_{3\b_2})(k_3\cdot \zeta_{4\b_4})~~.
\eeq

\section{D-brane action}
\label{ss:FT}
Let us compare these results to what one expects from the D-brane action in supergravity.  

As to the amplitude in Section \ref{ss:BC}, the relevant part of the D-brane action is (see
\eq{WZ})
\[
\frac{T_p}{\kappa}\int_{p+1}\hat{{\cal C}'}_{p-1}\wedge \hat B'~~.
\]
The amplitude calculated from this interaction thus becomes
\[
{\cal A}_{\rm sugra}=\frac{\kappa T_p}{(p-1)!} \epsilon^{\a_1\ldots\a_{p-1}\b_1\b_2} 
c_{\a_1\ldots\a_{p-1}}\zeta_{\b_1\b_2}~~.
\]

This is twice the string result. However, as suggested by R.~Russo, another diagram should be
considered in the supergravity computation. In fact, the bulk supergravity action given in the
appendix is not the whole story: the field strength of a $p+1$-form RR potential gets a correction
involving the $p-1$-form RR potential and the NS 2-form. This implies that there is a diagram in
which the D-brane emits an intermediate $p+1$-form potential. This contribution should be added to 
the contact amplitude given above. Checking the structure and the coefficient of this contribution
shows that this could explain the apparent mismatch.

For the amplitude in Section \ref{ss:RRC} the relevant part of the action is
\[
\frac{T_p}{\kappa}\int_{p+1}\hat{{\cal C}'}_{p-3}\wedge
\shalf\,\frac{(4\pi\a ')^2}{192\pi^2}\,\tr (\hat R\wedge\hat R)~~.
\]
Expanding $\hat R\wedge \hat R$ and writing everything in terms of canonically normalized fields
(see \eq{canG} and \eq{canC}) one obtains for the scattering amplitude in supergravity
\[
{\cal A}_{\rm sugra}=\frac{\sqrt{2}\,\pi^2\,\kappa^2\,T_p\a '^2}{6\,(p-3)!}\,
\epsilon^{\a_1\cdots\a_{p-3}\b_1\cdots\b_4}\,c_{\a_1\cdots\a_{p-3}}\,k_{3\b_1}\,k_{4\b_3}
(k_4\cdot\zeta_{3\b_2})(k_3\cdot\zeta_{4\b_4})~~.
\]

This time, the supergravity computation seems to be missing a factor of 2 relative to the string
result. Again, extra contributions might come from modifications of the RR field strengths
\cite{GHM}.
However, the details of this possible solution of the mismatch are left for future research.

\section{Orientifold planes} \label{ss:ORI}
The computations in the previous sections were done in oriented string theory. 
From the (oriented) type IIB theory, one can construct the (unoriented) type I theory
by modding out worldsheet parity.  
This operation makes another kind of objects enter the story: the
orientifold planes \cite{orientifold}. In Ref. \cite{Das} it was argued that
orientifold planes exhibit a coupling to gravity similar to the one we computed
for D-branes in Section \ref{ss:RRC}. Moreover, it was claimed that the orientifold 9-plane in
type I theory
couples with half the strength of the 32 type IIB D-branes (providing the Chan-Paton
factors \cite{PolWit}) together.

As remarked in Ref. \cite{Das}, the presence of that interaction can be checked by
computing an $RP^2$ diagram with three insertions. This is analogous to the
computation for D-branes: rather than cutting a hole in the sphere to obtain a
disk, one inserts a crosscap. For the case of all Neumann boundary conditions, the 
`crosscap state' representing the insertion
of a crosscap was discussed in Ref. \cite{callan}: apart from a different
normalization of $RP^2$ relative to disk amplitudes (giving an extra factor of 32), 
the crosscap operator is obtained by inserting $(-1)^n$ in every term of the exponent 
of the boundary state.

The computation of the $RP^2$ diagram is very similar to the one in Section
\ref{ss:RRC}. Because of the $(-1)^n$ insertions, every $z_i\bar z_j$
combination gets replaced by $-z_i\bar z_j$. In \eq{Int} this amounts to
replacing the 1 by $-1$. Analogous calculations to the ones which led to
\eq{Intresult} result in
\beq
{\cal I}_{\rm crosscap}=\frac{2\pi^4}{3}~~,
\eeq
half the result of the D-brane computation. Taking into account the factor of 32 from the crosscap
normalization, this reproduces precisely the coupling predicted in Ref. \cite{Das}.

\section{Discussion}
\label{ss:CONC}
By evaluating the appropriate string diagrams we have derived two specific terms of the
D-brane effective action. This nicely illustrates how superstring theory provides its own
consistency. If (intersecting) D-branes are to be included into the picture then from 
Ref.~\cite{GHM}
the terms in the effective D-brane action have to be there for the low energy
theory on the brane not to suffer from anomalies. 

Repeating the analysis for the orientifold, we have extended our results to the orientifold 
$p_1(R)$ coupling, first proposed in Ref.~\cite{Das}. Thus new evidence has been provided
for the relative factor $\frac 1 2$ between the D-brane and orientifold couplings.

Applying the above machinery one could in principle go beyond our results. The fourth order
curvature term can be treated analogously but may turn out to be a hard nut to crack, though. 

From a more general perspective one may ask the following question: is there room for a similar 
gravitational coupling in the worldvolume theory of the M-theory 5-brane
\cite{PST}? 
In Ref.~\cite{sen} Sen found such a coupling in the worldvolume action of the KK-monopole in
M-theory. It would be interesting to see whether his procedure can be applied to the M5-brane.

\medskip
\section*{Acknowledgments}
We would like to thank Marco Bill{\'{o}}, Andrea Pasquinucci,
Rodolfo Russo and Walter Troost for very helpful discussions.
This work was supported by the European Commission TMR programme ERBFMRX-CT96-0045.

\medskip
\section*{Note added}
While this paper was circulating as a preprint, the issue of gravitational couplings on the
M5-brane was addressed in Ref.~\cite{Mukhi}.

\noindent

\newpage

\appendix
\section{Conventions and normalizations}

We normalize the {\bf boundary state} as in \cite{billo9802}:
\begin{equation}
\label{bs3}
\ket{B,\eta}_{\rm R,NS} = {T_p\over 2}
\ket{B_X}\, \ket{B_\gh}\,{\ket{B_\psi,\eta}}_{\rm R,NS}
  \,{\ket{B_\sgh,\eta}}_{\rm R,NS}~~,
\end{equation}
where
\begin{equation}
\label{bs5}
\ket{B_X} = \delta^{(d_\bot)}(\hat q - y)
\exp\biggl[-\sum_{n=1}^\infty \frac{1}{n}\,
\a_{-n}\cdot S\cdot
\tilde \a_{-n}\biggr]\,
\ket{0;k=0}~~,
\end{equation}
\begin{equation}
\label{bs6}
\ket{B_\gh} = \exp\biggl[\sum_{n=1}^\infty
(c_{-n}\tilde b_{-n}
 - b_{-n} \tilde c_{-n})\biggr]\,{c_0 +\tilde c_0\over 2}\,\ket{q=1}
\,\ket{\tilde q = 1}~~, \end{equation}
and, in the NS sector in the $(-1,-1)$ picture,
\begin{equation}
\label{bs7}
\ket{B_\psi,\eta}_{\rm NS} = \exp\biggl[\ii\eta\sum_{m=1/2}^\infty
\psi_{-m}\cdot S \cdot \tilde \psi_{-m}\biggr]
\,\ket{0}~~,
\end{equation}
\begin{equation}
\label{bs8}
\ket{B_\sgh,\eta}_{\rm NS} =
\exp\biggl[\ii\eta\sum_{m=1/2}^\infty(\gamma_{-m}
\tilde\beta_{-m} - \beta_{-m}
  \tilde\gamma_{-m})\biggr]\,
  \ket{P=-1}\,\ket{\tilde P=-1}~,
\end{equation}
or, in the R sector in the $(-1/2,-3/2)$ picture,
\begin{equation}
\label{bs9}
\ket{B_\psi,\eta}_\R = \exp\biggl[\ii\eta\sum_{m=1}^\infty
\psi_{-m}\cdot S \cdot \tilde \psi_{-m}\biggr]
\,\ket{B_\psi,\eta}_\R^{(0)}~~,
\end{equation}
\begin{equation}
\label{bs10}
\ket{B_\sgh,\eta}_\R =
\exp\biggl[ \ii\eta\sum_{m=1}^\infty(\gamma_{-m}
\tilde\beta_{-m} - \beta_{-m}
\tilde\gamma_{-m})\biggr]\,
 \ket{B_\sgh,\eta}_\R^{(0)}~~,
\end{equation}
where, if we define
\begin{equation}
\label{bs14}
{\cal M}^{(\eta)} = C\Gamma^0\Gamma^{l_1}\ldots
\Gamma^{l_p} \,\left(
\frac{1+\ii\eta\Gamma_{11}}{1+\ii\eta}\right)~~,
\end{equation}
the zero mode parts of the boundary state are
\beq
\label{bsr0}
\ket{B_\psi,\eta}_\R^{(0)} =
{\cal M}_{AB}^{(\eta)}\,\ket{A} \ket{\tilde B}~~, 
\eeq
\beq
\label{bsrsg0}
\ket{B_\sgh,\eta}_\R^{(0)} =
\exp\left[\ii\eta\gamma_0\tilde\beta_0\right]\,
  \ket{P=-{1/ 2}}\,\ket{\tilde P=-{3/ 2}}~~.
\eeq
The matrix $S_{\mu \nu}$ is given by
\begin{equation}
S_{\mu \nu} = ( \eta_{\alpha \beta} , - \delta_{ij} )   ~~.
\label{smunu}
\end{equation}
The overall normalization factor $T_p$ is claimed to be fixed from factorization of
amplitudes of closed strings emitted from a disk \cite{frau, dv9707}. It is the tension of
the D$p$-brane
\begin{equation}
\label{tens}
T_p = \sqrt{\pi} (2\pi\sqrt{\alpha'})^{3 - p}~~.
\end{equation}
The GSO projection acts as follows:
\begin{equation}
\label{bs22ab}
\ket{B}_\NS
= {1\over 2} \Big( \ket{B,+}_\NS - \ket{B,-}_\NS \Big)~~,
\end{equation}
\beq
\label{bs22bb}
\ket{B}_\R  =
    {1\over 2} \Big( \ket{B,+}_\R + \ket{B,-}_\R\Big)~~.
\end{equation}

All this is explained in more detail in \cite{billo9802}. Conventions on the RR zero
modes and gamma matrices can be found in the Appendix of \cite{dv9707}. Here we only note that 
\beq
\left(\Gamma^\mu\right)^T = - C\,\Gamma^\mu\,C^{-1}  
\label{transp}
\eeq
and
\bea\label{convgamma}
\psi_0^\mu\, |A\rangle |{\widetilde B}\rangle
&=& \frac{1}{\sqrt{2}} \left(\Gamma^\mu\right)^A_{~C}
\,\left(\!\one\, \right)^B_{~D}|C\rangle\, |{\widetilde D}\rangle
\nonumber\\
{\widetilde \psi}_0^\mu \,|A\rangle |{\widetilde B}\rangle
&=& \frac{1}{\sqrt{2}} \left(\Gamma_{11}\right)^A_{~C}
\,\left(\Gamma^\mu\right)^B_{~D}\,
|C\rangle |{\widetilde D}\rangle~~.
\ena

In our computation we have put
\beq
\alpha '=2 ~~,
\eeq
so that the following OPEs hold \cite{friedan}:
\beqa
X^{\mu}(z)\,X^{\nu}(w)&=&-\eta^{\mu\nu}\log(z-w)~~,\\
\psi^{\mu}(z)\,\psi^{\nu}(w)&=&-\eta^{\mu\nu}(z-w)^{-1}~~,
\eeqa
where
\beq
X(z,\bar z)=X(z)+X(\bar z)~~.
\eeq
Expanding the fields in modes, this gives
\beqa
\comm{\a^{\mu}_m}{\a^{\nu}_n}&=&\eta^{\mu\nu}\,m\,\delta_{m+n}~~,\\
\comm{\a^{\mu}_0}{q^{\nu}}&=&-\ii\eta^{\mu\nu}~~,\\
\acomm{\psi^{\mu}_m}{\psi^{\nu}_n}&=&\eta^{\mu\nu}\,\delta_{m+n}~~,
\eeqa
where
\beqa
\ii\partial X^{\mu}(z)&=&\sum_n z^{-n-1}\, \a^{\mu}_n~~,\\
\ii\psi^{\mu}(z)&=&\sum_r z^{-r-\frac{1}{2}}\,\psi^{\mu}_r~~.
\eeqa

For the {\bf graviton vertex operator in the (-1,-1) picture} we take
\beq
V_g^{(-1)}=\zeta_{\mu\nu}\, \psi^{\mu}\,\tilde\psi^{\nu}\,e^{\ii k\cdot X}~~.
\eeq
This normalization is such that
\beq
\braket{V_g^{(-1)}}{V_g^{(-1)}}=-\tr \zeta^2~~.
\eeq
It can be  checked \cite{dv9707} that, with this normalization,
\beq
\bra{0}V_g^{(-1)}\ket{B}_{\NS}
\eeq
reproduces the one point function of the canonically normalized graviton in supergravity, where
the D-brane is described by the DBI action. The canonically normalized graviton field is obtained
by expanding
\beq
g_{\mu\nu}=\eta_{\mu\nu}+2\kappa\, h_{\mu\nu}~~,
\eeq
(where $\kappa=8\pi^{7/2}\a '^2g_s=\sqrt{8\pi G_N}$)
in the bulk supergravity action in Einstein frame
\begin{eqnarray}
S_{\rm IIA,B}  &=& -{1\over 2\kappa^2}
 \int d^{10}x \sqrt{-g}\ \Bigl[
 R + {1\over 2} (d\Phi)^2 
 +{1\over 12} e^{-\Phi} (dB')^2 \nonumber \\
&&\ \ \ \  + \sum  {1\over 2(p+2)!}
e^{(3-p)\Phi/2} (d {\cal C}'_{p+1})^2 \ \Bigr]
\end{eqnarray}
(see, for instance \cite{bachas}).
The DBI action is given in string frame ($G_{\mu\nu}=e^{\Phi/2}\, g_{\mu\nu}$) by
\beq
S_{BI}=-\frac{T_p}{\kappa}\int d^{p+1}\xi\,e^{-\Phi}\sqrt{-\det [\hat{G}_{\a\b}+\hat{B'}_{\a\b}
+2\pi\a '\,F_{\a\b}]}~~.
\eeq
(Here $\hat G$ denotes the pullback to the D~brane worldvolume of the bulk field $G$ etc.)
In addition, the D-brane action contains a WZ part
\beq \label{WZ}
S_{WZ}=\frac{T_p}{\kappa}\int_{p+1}\hat{{\cal C}'}\wedge e^{2\pi\a '\,F+\hat{B'}}\wedge\sqrt{\hat{A}(\hat
R)}~~,
\eeq
where
\beq
\hat A(\hat R)=1+\frac{(4\pi\a ')^2}{192\pi^2}\tr (\hat R\wedge\hat R)+\ldots~~.
\eeq
These actions can be written in terms of canonically normalized bulk fields by substituting
\beqa
G&=&e^{\Phi/2}(\eta_{\mu\nu}+2\kappa\, h_{\mu\nu})\label{canG}\\
\Phi&=&\sqrt{2}\kappa\,\chi\\
B'&=&\sqrt{2}\kappa\, e^{\Phi/2} B\\
{\cal C}'&=&\sqrt{2}\kappa\, {\cal C}~~.\label{canC}
\eeqa

For the normalization of the {\bf graviton vertex operator in the (0,0) picture}, we follow
Section~3.4 of \cite{gsw}:
\beq
V_g^{(0)}=\frac{\kappa}{\pi}\,\zeta_{\mu\nu}\,(\partial X^{\mu}+\ii \,k\cdot\psi\,\psi^{\mu})
(\bar\partial
X^{\nu}+\ii\,k\cdot\tilde\psi\,\tilde\psi^{\nu})\,e^{\ii k\cdot X}
\eeq
(where we put $\a '=2$ rather than $\a '=1/2$ as in \cite{gsw}).

For the {\bf Kalb-Ramond field} we would correspondingly find
\beq \label{KR}
V_B^{(0)}=\frac{\kappa}{\sqrt{2}\pi}\,\zeta_{\mu\nu}\,(\partial X^{\mu}+\ii\,k\cdot\psi\,
\psi^{\mu})(\bar\partial
X^{\nu}+\ii\,k\cdot\tilde\psi\,\tilde\psi^{\nu})\,e^{\ii k\cdot X}~~.
\eeq
(The factor of $1/\sqrt{2}$ is due to the different rescaling between the canonically
normalized fields and the fields appearing in the string action. The $\zeta_{\mu\nu}$ in the
vertex operators correspond to polarizations of canonically normalized fields.)

Finally, we introduce a (canonically normalized) {\bf out-state for a RR-potential}
\cite{billo9802}:
\beq
\bra{{\cal C}_{m+1}}=\shalf (\bra{{\cal C}_{m+1},+}+\bra{{\cal C}_{m+1},-})
\eeq
with
\beq
\bra{{\cal C}_{m+1},\eta}=e^{i\eta \b_0\tilde\c_0}\N_{AB}{}_{-3/2}\bra{A}\,{}_{-1/2}\bra{\tilde B}~~,
\eeq
where we defined
\beq
\N_{AB}=\frac{(C\Gamma^{\mu_1\cdots\mu_{m+1}})_{AB}}{2\sqrt{2}\,(m+1)!}\,
c_{\mu_1\cdots\mu_{m+1}}~~.
\eeq
(Here $c$ denotes the polarization tensor of ${\cal C}$). 
The factor $\frac{1}{2\sqrt{2}}$ is justified by computing
\beq
\braket{{\cal C}_{m+1}}{{\cal C}_{m+1}}=\frac{1}{(m+1)!}\,c_{\mu_1\cdots\mu_{m+1}}\, c^{\mu_1\cdots\mu_{m+1}}~~.
\eeq
As a check, one can derive \cite{marco} the $e^{2\pi\a 'F}$ part of the WZ term of the
D-brane action by projecting the states $\bra{{\cal C}_{p+1-2n}}, n=0,1,\ldots$ on
the boundary state with $F$ turned on.

So far we have only computed one-point functions in the presence of a D-brane. The ghost zero
modes have been taken into account by not integrating over the position of the inserted vertex
operator. Because of the peculiar ghost structure of the boundary state, it is not so clear how
the ghosts should be treated when one inserts more than one vertex operator (i.e. which part of
the $SL(2)$ group volume still has to be divided out). The procedure we will
follow (and motivate below) consists of integrating over the positions of 
all vertex operators except the one inserted at infinity.

To motivate this procedure, we compute the two graviton amplitude for both
polarizations along the brane (up to powers of $\ii$):
\beq
\bra{0}V_g^{(-1)}\,V_g^{(0)}\ket{B}_{\NS}=\kappa\, T_p\, q^2
\,\tr(\zeta_1\cdot\zeta_2)\,B(-\frac{t}{2},2q^2+1)~~,
\eeq
where $q^2=\shalf(k_2\cdot S\cdot k_2)$ and $t=-(k_1+k_2)^2$. 
This is the correct result (see for instance \cite{gm}).

\end{document}